\documentclass[prl,aps,twocolumn]{revtex4}
\usepackage{amssymb}
\usepackage{amsmath}
\usepackage{graphicx}
\usepackage{t1enc}
\usepackage[utf8]{inputenc}

\begin{document}

\title{Quantum accelerometer: distinguishing inertial Bob from his accelerated twin Rob by a local measurement}

\author{Andrzej Dragan}
\address{Institute of Theoretical Physics, University of Warsaw, Ho\.{z}a 69, 00-049 Warsaw, Poland}
\author{Ivette Fuentes\footnote{Previously known as Fuentes-Guridi and Fuentes-Schuller.}}
\address{School of Mathematical Sciences, University of Nottingham, Nottingham NG7 2RD, United Kingdom}
\author{Jorma Louko}
\address{School of Mathematical Sciences, University of Nottingham, Nottingham NG7 2RD, United Kingdom}

\begin{abstract}
A single quantum system, such as Unruh-DeWitt detector, can be used to determine absolute acceleration by local measurements on a quantum field. To show this, we consider two kinematically indistinguishable scenarios:  an inertial observer, Bob, measuring the field of an uniformly accelerated cavity, and his non-inertial twin Rob accelerating and making measurements in a stationary cavity. We find that these scenarios can be distinguished in the non-relativistic regime only by measurements on highly excited massive fields, allowing one to detect non-inertialness of the reference frame.
\end{abstract}

\maketitle

A possible way to determine whether a given frame of reference is inertial or not is by using two stationary and initially synchronized ideal clocks at a fixed distance from each other. If the clocks desynchronize - the frame is inevitably non-inertial~\cite{note:refframe}. In this paper we show that it is possible to distinguish whether a reference frame is inertial or not using a single point-like quantum system which measures the states of a quantum field locally. The measurement outcome determines the system's absolute acceleration.

It is well known that uniformly accelerated observers moving in a Minkowski vacuum perceive a thermal bath with a temperature proportional to their acceleration \cite{Unruh}. However, the observation of the thermal bath is not enough to claim that the frame of reference in which the observations are carried out is non-inertial. Detection of thermal particles could equally well indicate that the frame is inertial, but the space is filled with thermal radiation. Even when the accelerated Rob detects thermal radiation and his inertial twin, Bob, claims that the field is in the vacuum state, is not enough to ensure that Rob is indeed accelerated. In order to claim so, one has to consider as well the opposite situation. The state of the field from Bob's perspective must be determined, when Rob claims that the field is in the vacuum state. Only when a difference appears between these scenarios it is possible to construct a quantum accelerometer. 

\begin{figure}
\includegraphics[width=7cm]{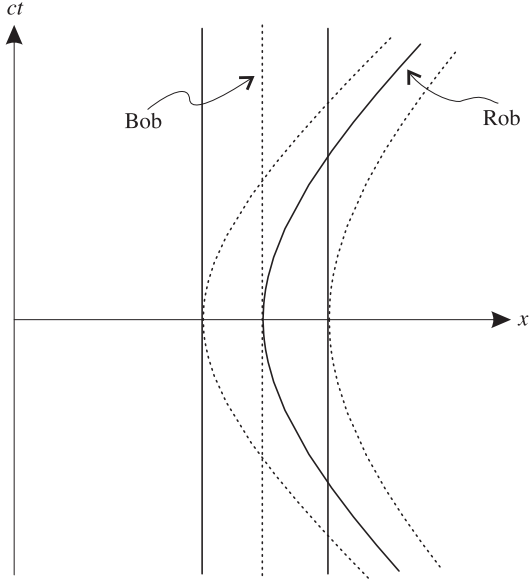}
\caption{\label{scheme}Inertial Bob moving through an accelerated Rob's cavity (dashed lines) and Non-inertial Rob accelerating through stationary Bob's cavity (solid lines).}
\end{figure}

In this paper we analyze and compare both situations considering real Klein-Gordon massive fields confined in finite cavities and being measured by single point-like Unruh-DeWitt detectors \cite{Unruh, DeWitt, Crispino}. Similar settings have been recently considered to generate maximum entanglement between cavity field modes \cite{Downes}. The scenarios we consider are carefully constructed in such a way that they are possibly indistinguishable from a kinematical point of view. We consider inertial Bob moving through Rob's cavity in uniform acceleration while accelerated Rob moves through Bob's stationary cavity - see Fig.~\ref{scheme}. We compare the probability of detector excitation in both scenarios. When the probabilities differ, it is possible to not only determine who is non-inertial but also to estimate the absolute acceleration. 

In \cite{VerSteeg} the entanglement generated between two point-like detectors is used to distinguish between an expanding (de~Sitter) spacetime and a flat  spacetime. In this scheme, a single detector is not capable of distinguishing in which spacetime the motion takes place. The reason for this is that the response of the detector in the de~Sitter vacuum is identical to the response of the detector in a thermal ensemble of fields in flat spacetime \cite{VerSteeg,Gibbons}. Interestingly, we show that a single point-like detector can be used to determine whether its trajectory in flat spacetime is inertial or not. It is not necessary to use non-local quantum properties, such as entanglement, to make this distinction. A~local measurement on the field is shown to be enough.

Let us consider an Unruh-DeWitt detector which Hamiltonian 
\begin{equation}
\label{UDW-detector}
\hat{H}_I(\tau) \propto \epsilon(\tau)\hat{\phi}\left[x(\tau) \right]\left(\hat{d}e^{-i\omega\tau} + \hat{d}^\dagger e^{i\omega\tau} \right)
\end{equation}
describes the interaction between a real scalar field $\hat{\phi}$ of mass $m$ and a point-like particle (detector) characterized by an annihilation operator $\hat{d}$.  $x(\tau)$ is the trajectory of the detector as a function of its proper time $\tau$, $\omega$ is the frequency between the detector's ground and excited states and the window function $\epsilon(\tau)$ turns the Hamiltonian on and off~\cite{note:window}. 
The field  operator in Minkowski coordinates $\hat{\phi}(t, x)$ is given by
\begin{equation}
\hat{\phi}(t, x) = \sum_{k=1}^\infty F_k(x)\left(e^{-i\omega_k t}\hat{a}_k + e^{i\omega_k t}\hat{a}^\dagger_k\right)
\end{equation} 
where $F_k(x)$ are normalized modes of the real massive Klein-Gordon equation satisfying appropriate boundary conditions, and $\hat{a}_k$ are associated annihilation operators.  We consider the detector to be initially in the ground state $|g\rangle$ and the field in the Fock state $|n_1\rangle$ of the lowest-energy mode. To first order in a perturbative expansion, the probability amplitude for the atom to undergo a transition to the excited state $|e\rangle$ and the field to jump to an arbitrary state $|\psi\rangle$, is given by
\begin{equation}
\label{amplitude}
{\cal A}_\psi = -i\int_{-\infty}^\infty \mbox{d}\tau \langle e|\langle\psi| \hat{H}_I(\tau) |g\rangle |n_1\rangle.
\end{equation}
Therefore, the probability ${\cal P} = \sum_\psi |{\cal A}_\psi|^2$ for the detector to click  yields
\begin{eqnarray}
\label{probability}
{\cal P} &\propto& \sum_{k=1}^\infty\left|\int\text{d}\tau\, \epsilon(\tau)F_k[x(\tau)]e^{i\left(\omega\tau+\omega_k t(\tau)\right)}\right|^2\nonumber \\
& &+ n_1\left|\int\text{d}\tau\, \epsilon(\tau)F_1[x(\tau)]e^{i\left(\omega\tau+\omega_1 t(\tau)\right)}\right|^2\nonumber \\ 
& &+ n_1\left|\int\text{d}\tau\, \epsilon(\tau)F_1[x(\tau)]e^{i\left(\omega\tau-\omega_1 t(\tau)\right)}\right|^2.
\end{eqnarray}

\begin{figure}
\includegraphics[width=\columnwidth]{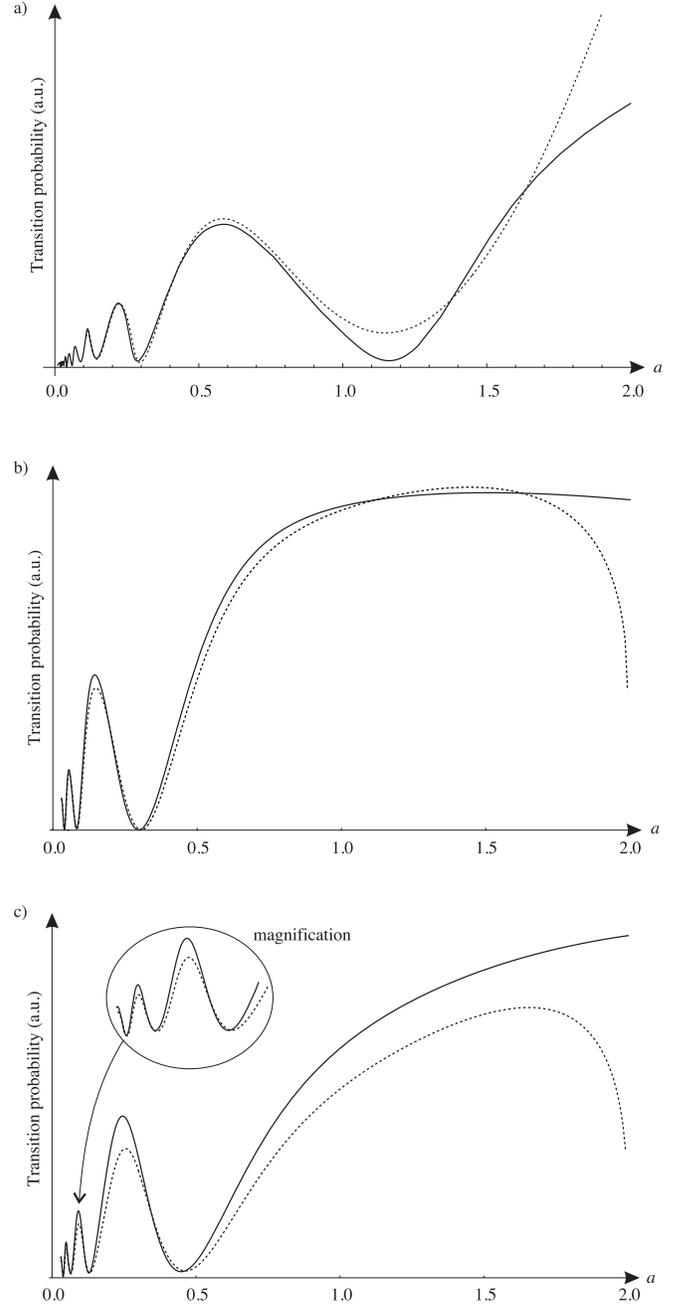}
\caption{\label{plots}Probability of a transition of Bob's detector (dashed line) and Rob's detector (solid line) as a function of acceleration: a) for the vacuum state of the field of the mass $m=0.2$; b) for a highly populated ground state of the cavity and mass $m=0.2$; c) for a highly populated ground state of the cavity and mass $m=2$.}
\end{figure}

In the case where the cavity is stationary and the field vanishes at $x=\pm\frac{L}{2}$, the normalized field modes are given by
\begin{equation}
F_k(x)=\frac{1}{\sqrt{k\pi}}\sin\left( k\pi\frac{x+L/2}{L}  \right)
\end{equation} 
with corresponding frequencies $\omega_k = \sqrt{\left(\frac{k\pi}{L}\right)^2+m^2}$. We consider Rob and his detector moving along an uniformly accelerated trajectory $x(\tau) = -a^{-1}\left[ \cosh (a\tau) - 1\right]$, so that the relation between Bob's time $t$ and Rob's proper time $\tau$ is given by: $t(\tau) = a^{-1}\sinh (a\tau)$. We also choose $\epsilon(\tau)$ such that the interaction is on, $\epsilon = 1$, only while the atom is inside the cavity, otherwise $\epsilon = 0$. In this case, the total probability ${\cal P_{\text{Rob}}}$ for Rob's detector transition \eqref{probability} involves the following limits of  integration: $\pm\frac{c}{a}\text{arccosh}\left(1+\frac{aL}{2c^2}\right)$.

We have numerically studied the probability ${\cal P_{\text{Rob}}}$ as a function of the acceleration for various settings, letting $L=1$. The case when the cavity is in the vacuum state ($n_1=0$) and the field mass is small ($m=0.2$) is represented by a solid line in the Fig.~\ref{plots}a. For practical purposes, it was sufficient to truncate the series involved in calculating \eqref{probability} at a finite $k$,  in this case being $15$. When the mode of the cavity is strongly occupied, $n_1\gg 1$, one can neglect the contribution from the first-term in the series \eqref{probability} and consider only the last two terms. In this case, the probability ${\cal P_{\text{Rob}}}$ changes its behavior, as shown with solid lines in  Fig.~\ref{plots}b for $m=0.2$ and Fig.~\ref{plots}c for $m=2$. 

In the second scenario the cavity is uniformly accelerated and Bob and his detector are moving freely through the cavity. The coordinate transformation to the cavity's rest frame is given by the Rindler transformation
\begin{equation}
\label{Rindlertransformation}
\tau = a^{-1}\text{atanh}\frac{t}{x}, ~~~\chi = \sqrt{x^2-t^2}.
\end{equation}
In these coordinates the field vanishes again at the boundaries of the cavity at $\chi_{1,2} = \frac{c^2}{a}\pm\frac{L}{2}$ and the Klein-Gordon equation has the form
\begin{equation}
\left[\frac{1}{a^2\chi^2}\frac{\partial^2}{\partial\tau^2}-\frac{\partial^2}{\partial\chi^2}-\frac{1}{\chi}\frac{\partial}{\partial\chi} +m^2 \right]\hat{\phi} = 0.
\end{equation}
In this case the cavity modes in the coordinates $(\tau, \chi)$ are given by
 \begin{equation}
 F_k(\chi) = N_k\left[I_{-i\Omega_k}(m\chi_1)I_{i\Omega_k}(m\chi)-I_{i\Omega_k}(m\chi_1)I_{-i\Omega_k}(m\chi)  \right]\nonumber
 \end{equation}
where $I_{\alpha}(z)$ is the modified Bessel function of the first kind, $\Omega_k$ are frequencies such that the expression in the square brackets for $\chi=\chi_2$ vanishes, and $N_k$ are normalization constants.  In the massless case, the Klein-Gordon equation is conformally-invariant and the field modes in the accelerated frame have the same form as the inertial modes.  Bob's free trajectory according to Rob accelerating with the cavity is given by $\chi(t) = \sqrt{a^{-2}-t^2}$ for $t<a^{-1}$, i. e. before the detector falls onto Rob's event horizon, and relation between the Rob Rindler's time $\tau$ and Bob Minkowski's time $t$ from Rob's perspective is given by $\tau(t)=a^{-1}\text{atanh}(at)$. We obviously assume that the whole cavity is above the event horizon, which boils down to the assumption that $L<2a^{-1}$. If we choose $\epsilon(t)$ again such that the interaction is on when Bob's detector is inside the cavity, the limits of integration in the equation \eqref{probability} defining the probability ${\cal P}_{\text{Bob}}$ of Bob's detector to click effectively become $\pm a^{-1}\sqrt{aL\left(1-aL/4\right)}$.   

In order to study the probability  ${\cal P}_{\text{Bob}}$ of Bob's detector to click, we have numerically evaluated the frequencies $\Omega_k$ and normalization constants $N_k$ as functions of acceleration $a$. The case in which the field is in the vacuum ($n_1=0$) and the mass is small ($m=0.2$) in this scenario is represented by a dashed line in the Fig.~\ref{plots}a. We see that the probability of detector excitation as a function of the acceleration differs in both scenarios only in the limit of relativistic accelerations, $a\sim L^{-1}$. This is exactly what should be expected in the limit of $m\ll 1$, when the Klein-Gordon equation becomes conformally invariant. In this case, the field operator in the coordinates $(\tau, \xi)$, where $\xi=a^{-1}\log{a\chi}$, takes exactly the same form as in the coordinates $(t, x)$, and the interaction Hamiltonian \eqref{UDW-detector} coincides in the two scenarios up to the kinematical difference between the trajectories $x(\tau)$, and $\xi(t)$. The discrepancy is related to an asymmetry of Rob's and Bob's observations of the mutual trajectories when the velocities become relativistic. We conclude from the previous analysis that is not possible to determine in our scenario if a reference frame is inertial or not using local measurements on the field if the field is massless or, in the massive case, if the field is in the vacuum state.

The same argument can be used to explain the similarity between the plots of ${\cal P}_{\text{Bob}}$ (dashed line) and ${\cal P}_{\text{Rob}}$ (solid line) in the Fig.~\ref{plots}b drawn for small mass $m=0.2$ and high excitation of the cavity field $n_1\gg 1$. Again, approximate conformal invariance suppresses all the differences for small accelerations. Surprisingly, the situation is radically different when considering massive fields in highly populated field mode states. Interesting effects arise when we depart from the regime of small mass. The analysis carried out for $m=2$ shows that the two scenarios can be distinguished even for small accelerations, as shown in the Fig.~\ref{plots}c. The fact that in the case of high mass and no cavity excitations there are no observable differences between ${\cal P}_{\text{Bob}}$ and ${\cal P}_{\text{Rob}}$ in the limit of small accelerations,  suggests that the presence of massive particles in the system plays a crucial role in the local detection of absolute acceleration.

Point-like quantum systems moving in spacetime can be used as accelerometers determining if a reference frame is inertial or not. To demonstrate this, we propose a scheme which involves an Unruh-DeWitt detector making measurements on a cavity field. We find that it is not possible to measure absolute acceleration in the case where the cavity field is massless. However, the presence of the field mass allows for the measurement of the absolute acceleration. It is the fact that the mass breaks the conformal invariance of the field equation that makes our scheme possible. Interestingly, the presence of gravitational mass always leads to space-time curvature and consequently to the impossibility of defining a globally inertial frame. Therefore, it has been interesting to find that the presence of mass is also necessary to detect locally the absolute acceleration, within the quantum picture.

\section{Acknowledgements}
We would like to thank R.~B. Mann, D.~Sudarsky, E.~Mart\'{i}n-Mart\'{i}nez, and K. Pachucki for interesting discussions and useful comments.  I.~F. thanks EPSRC  [CAF Grant EP/G00496X/2] for financial support. 
J.~L. was supported in part by STFC (UK) Rolling Grant PP/D507358/1. 

\bibliographystyle{apsrev}
\bibliography{references}

\begin{thebibliography}{31}
\bibitem{note:refframe}
Throughout this work we use the term ``frame of reference'' as a special case of a coordinate system in which a set of labeled stationary ideal clocks and rigid rods can be used to determine space-time coordinates. Both in Minkowski and Rindler coordinates such construction is always possible.
\end{thebibliography}

\end{document}